\documentclass[12pt]{article}
\usepackage{amsmath}
\usepackage{amssymb}
\usepackage{graphicx}
\usepackage{epsfig}
\usepackage{color}
\usepackage{fullpage}

\newcommand\pubnumber{SLAC-PUB-12930}
\newcommand\pubdate{October 30, 2007}

 \def\SLAC{Stanford Linear Accelerator Center\\
  2575 Sand Hill Road, Menlo Park, California, 94025, USA}
\def\STANKI{Dept. of Physics and Kavli Institute for Particle Astrophysics and Cosmology\\
 Stanford University, Stanford, CA 94305-4060}
\def\doeack{\footnote{Work supported by the US Department of Energy,
                     contract DE--AC02--76SF00515.}}

\def\Title#1{\begin{center} {\Large #1 } \end{center}}
\def\Author#1{\begin{center}{ \sc #1} \end{center}}
\def\Address#1{\begin{center}{ \it #1} \end{center}}
\def\andauth{\begin{center}{and} \end{center}}

\newcommand\pubblock{\rightline{\begin{tabular}{l} \pubnumber\\
         \pubdate \end{tabular}}}

\newcommand{\cs}{\rm cm^3s^{-1}}
\newcommand{\so}{(\sigma v)_0}
\newcommand{\sv}{(\sigma v)_1}

\title{Relativistic dark matter at the galactic center}
\author{Mustafa A. Amin \&  Tommer Wizansky}

\date{ }
\begin{document}

\begin{titlepage}
\pubblock

\vfill \Title{Relativistic dark matter at the galactic center}
\vfill \Author{Mustafa A. Amin\footnote{email: mamin@stanford.edu}} \Address{\STANKI}  \andauth
\Author{Tommer Wizansky\footnote{email: twizansk@stanford.edu}$^{,}$\doeack}\Address{\SLAC}

\begin{abstract}
In a large region of the supersymmetry parameter space, the annihilation cross section for neutralino dark matter is strongly dependent on the relative velocity
of the incoming particles. We explore the consequences of this velocity dependence in the context of indirect detection of dark matter from the galactic center. We find that the increase in the annihilation cross section at high velocities leads to a flattening of the halo density profile near the galactic center and an enhancement of the annihilation signal.
\end{abstract}

\vfill
\end{titlepage}
\def\thefootnote{\fnsymbol{footnote}}
\setcounter{footnote}{0}

\begin{section}{Introduction}
Indirect detection is one of the most promising avenues for the discovery of
dark matter through its non-gravitational effects.  Many efforts are underway and more are planned to detect the products of dark matter annihilations  \cite{GLAST,VERITAS,HESS,MAGIC,EGRET,Barwick:1997tv,PAMELA,BESSPolar,INTEGRAL,AGILE}.  The best places to look for the annihilation signal are regions where the density of dark matter is expected to be high, for example, centers of dark matter halos \cite{Ullio:2002pj}, center of stars \cite{Salati:1989, Moskalenko:2007ak} and neighborhoods of compact objects \cite{Bertone:2007ae}.

We concentrate on the sub-parsec region around the super-massive black hole (SBH) at  center of our galaxy ($M_{\rm bh}\approx4\times 10^6 M_{\odot}$ \cite{Genzel:2003cn,Schodel:2003gy}). Gondolo and Silk \cite{Gondolo:1999ef} argued that a sharp dark matter spike should form around the SBH leading to a large enhancement of the annihilation signal. Subsequent authors (for example \cite{Ullio:2001fb,Merritt:2003qk, Bertone:2005hw}) qualified this statement, pointing out several phenomena which would have the effect of smoothing and reducing the spike.  The debate over the existence of a dark matter spike at the center of the galaxy has yet to be resolved. For the purpose of this paper, we assume that a spike does exist. 

In this study we discuss a new correction to the predictions for the annihilation rate and halo profile around
the SBH.  We point out that near the black hole the dark matter particles will be moving
sub-relativistically ($v/c\lesssim0.2$). This is in contrast to the usual assumption whereby the dark matter is taken
to be cold and slow.  In fact, most previous calculations (see for example \cite{Gondolo:1999ef}) have been performed in the limit $(v/c)\rightarrow0$ where
$v$ is the relative velocity between particles.  For a certain class of supersymmetric dark matter
models, the cross section for annihilation can be enhanced by several orders of magnitude in the vicinity of the SBH due a strong dependence on $v$.  In the presence of a central dark matter spike this can produce a measurable correction to the observed annihilation signal.  In addition, the enhanced cross section leads to depletion of the spike and a widening of the ``annihilation core''. We explore these two effects for a variety of spike profiles to account for the many astrophysical uncertainties regarding the nature of the density profile. 

We find that the enhancements in the annihilation signals occur primarily
in models for which the indirect detection signals are too small to be 
seen by current experiments.   However,
these models are quite plausible theoretically and are even preferred by
some criteria.  We can easily imagine a scenario in which particle physics
experiments point to one of these theories as a correct description of 
nature.  This will motivate dedicated 
 gamma ray observations
concentrating on objects where dark matter is likely to be concentrated.
We will argue that, in this situation, the velocity-dependent enhancement 
of the annihilation cross section must be taken into account.

The rest of the paper is organized as follows.  In Section \ref{pp} we give a brief review of supersymmetric dark matter
and enumerate the circumstances whereby a strong enhancement to the annihilation cross section may arise.
In Section \ref{ap} we estimate the corrections to the halo profile arising from the
enhanced annihilation rate and calculate corrections to the annihilation signal.  Our conclusions are presented in Section \ref{disc}.

\end{section}
\begin{section}{Supersymmetric dark matter}
\label{pp}
For the purpose of this study we restrict ourselves to the minimal
supersymmetric standard model (MSSM).   In this class of theories there exist
four neutral fermionic mass eigenstates -- the neutralinos.  The lightest
of these is often the lightest superpartner in
the theory (LSP) and provides a good dark matter candidate. We are interested in describing the conditions under which the annihilation of the LSP to standard model particles exhibits a strong velocity dependence leading to an enhancement of
the indirect detection signal. 

A sample of the most important Feynman diagrams contributing
to neutralino annihilations
are depicted in Figure \ref{figDiags}. First, a pair of neutralinos may exchange a fermion
superpartner (sfermion), producing two standard model fermions. Fermions may also be produced
through an s-channel exchange of a heavy scalar, in this case the $A^0$ Higgs boson.
Notably, this diagram does not admit a p-wave component, a fact which will be important
in the coming analysis. Finally, the neutralinos may annihilate to standard model gauge bosons. In Figure
\ref{figDiags} we present the annihilation to two $Z^0$ bosons via the exchange of
a heavier neutralino.
\begin{figure}[t]
\centering
\includegraphics{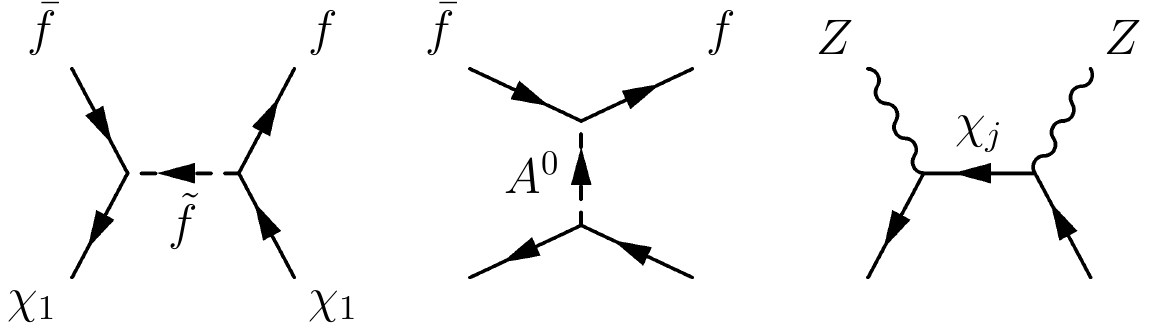} \caption{A typical set of Feynman diagrams
contributing to the self annihilations of a
  neutralino into Standard Model particles. The LSP is denoted by $\chi_1$, $\chi_j$
  is a heavier neutralino and $A^0$ is the neutral CP-odd Higgs boson.}
\label{figDiags}
\end{figure}

In the MSSM, neutralinos are Majorana particles.  This leads to a 
well-known helicity
suppression of the amplitude for pair annihilation into light
fermions~\cite{Goldberg}.
If $\chi$ denotes the dark matter particle and $f$ the fermion, the s-wave
cross section
will be suppressed by a factor of 
$$         z^2 = m_f^2/m_\chi^2   \ . $$
For the annihilation of a 200 GeV neutralino to Standard Model leptons,
$z^2$ is less than  $10^{-4}$.  Consequently, the p-wave annihiliation,
which is suppressed only by $v^2/c^2$, 
may dominate. We can therefore conclude that for
models where the LSP annihilates primarily to fermions, the
annihilation cross section will exhibit a strong velocity
dependence. It is this effect which lies at the core of our present
work. In the next few paragraphs, we review how these helicity-suppressed cross sections arise in
the MSSM parameter space. A more complete description can be found, for example, in
\cite{Griest:1988ma}. For an excellent review of the MSSM and
supersymmetry in general see \cite{Martin:1997ns}.

As mentioned, only annihilations to fermions undergo
helicity suppression. We would like to identify the regions of parameter space for which
the dominant annihilation channels do undergo helicity suppression and the resulting process is
p-wave.  It is this class of models which will exhibit a strong velocity dependence.

In the MSSM, each neutralino is a linear combination of the superpartners of two neutral
gauge bosons and two neutral Higgs bosons. It is typically parametrized by
$$
\chi_i = Z_{i1}\tilde{B}^0+Z_{i2}\tilde{W}^0+
Z_{i3}\tilde{H}_1^0+Z_{i4}\tilde{H}_2^0,
$$
where $\chi_i$ is the $i^{th}$ neutralino and tildes denote superpartners.
The partners of the $B^0$, $W^0$ and Higgs bosons are usually called wino,
bino and Higgsino respectively. Of these four only the bino is a gauge singlet, meaning
that it does not interact with gauge bosons. Thus, by making $Z_{11}$ large
compared to the other components, we can eliminate the third diagram in Figure \ref{figDiags},
leaving only fermionic processes. The annihilation
of Majorana particles through a scalar coupling can only take place in the
s-wave.  Thus, if the second diagram were to dominate over the first, the cross
section would indeed be helicity suppressed but no strong velocity
dependence would arise.  To suppress this diagram we demand
that the $A^0$ boson is significantly heavier than the fermion superpartners.
We must also make sure that no resonance enhances the $A^0$
diagram, that is, $m_{A}$ cannot be too close to $2m_{\chi}$.

We may now ask how generic are these constraints? The condition of
large $Z_{11}$ is quite generic.  The theoretically compelling
assumption of gauge unification naturally leads to a bino that is
lighter than the wino by a factor of two \cite{Inoue:1982pi}.  In many classes of models, for example, minimal supergravity, the condition of electroweak symmetry breaking requires that the higgsinos are quite heavy. In these cases the LSP is  mostly bino. The mass of
the $A^0$ boson is, in principal, unconstrained and can easily be
large enough to suppress the $A^0$ exchange diagram.

\begin{figure}[t]
  \centering
  \begin{tabular}{cc}
  \includegraphics[width=3in]{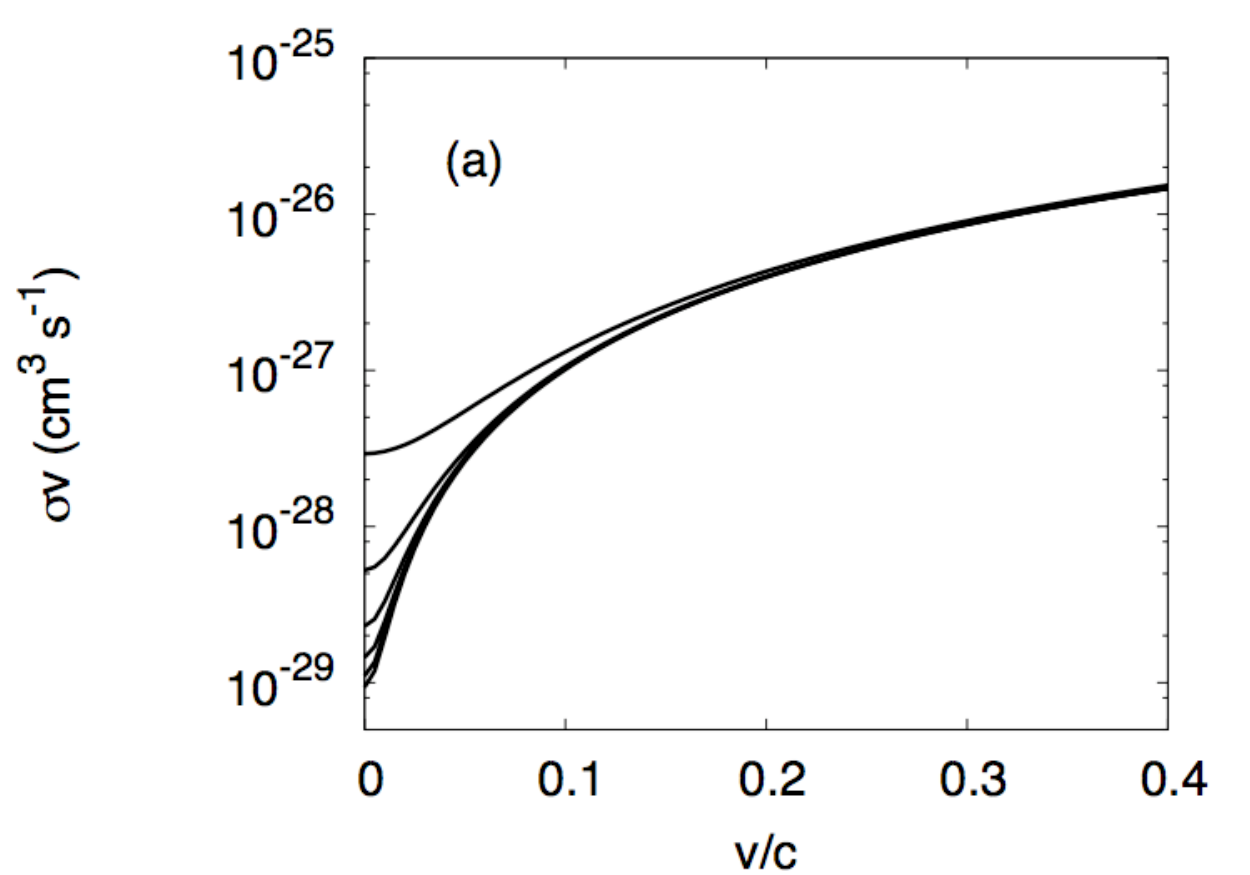} &
  \includegraphics[width=3in]{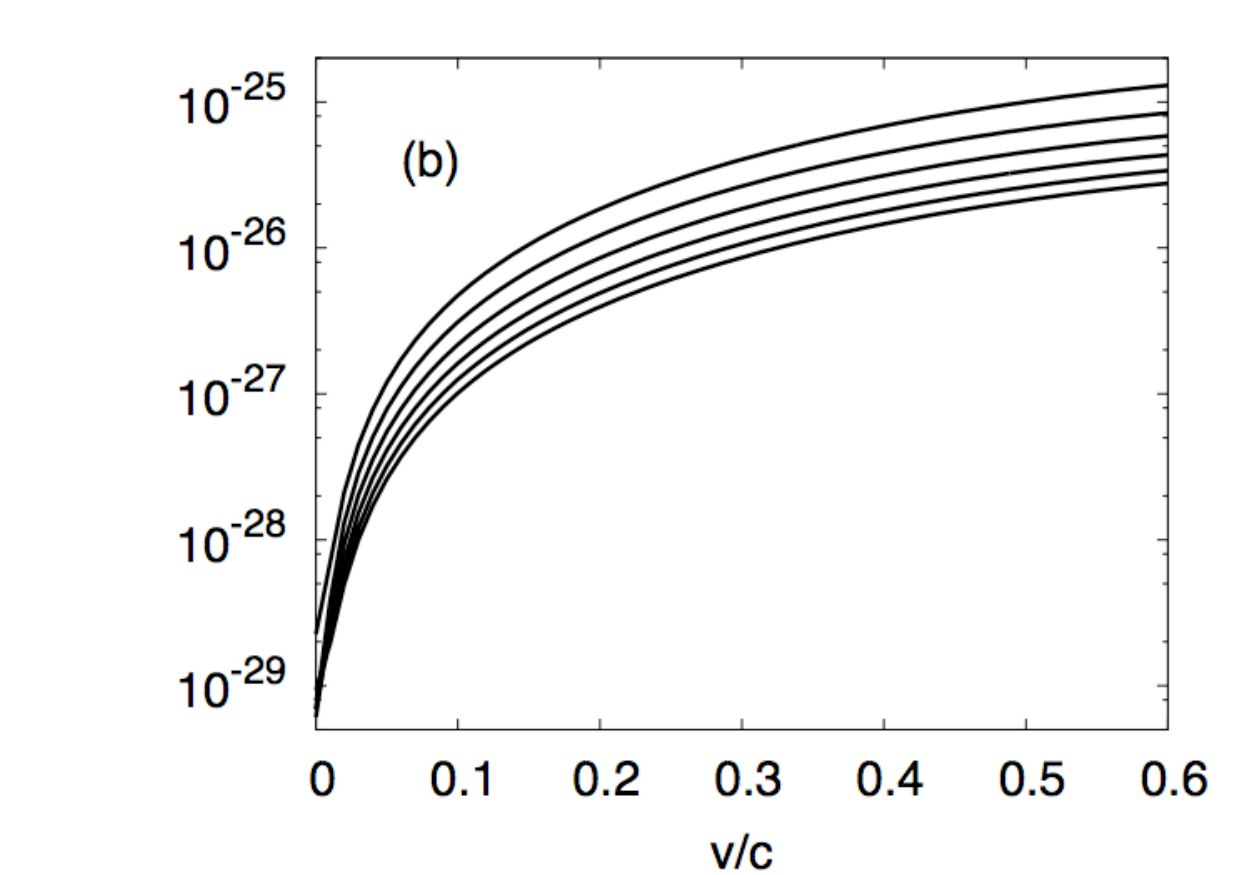}
  \end{tabular}
  \caption{$\sigma v$ as a function of $(v/c)$ for sample models from the $\tilde\tau$ coannihilation
    region. In  (a) $m_{A}$ is scanned, increasing from top to bottom, and in (b) $m_{\tilde\tau}$ is scanned, increasing from top to bottom.}
  \label{figSigvHelicity}
\end{figure}

Thus, helicity-suppressed dark matter annihilation is quite likely in 
models of supersymmetry.  This implies a strong dependence of the
annihilation rate  on the relative velocity of the incoming particles.  In
the following, we will consider sample MSSM models from this region of 
parameter space.  We will choose models for which the predicted dark
matter relic density
agrees with value $\Omega_\chi h^2 \approx 0.1$ given by the WMAP
experiment~\cite{Spergel:2006hy}.
All cross sections and relic density calculations were performed using the
DarkSUSY software package~\cite{Gondolo:2004sc}.

In Figure \ref{figSigvHelicity},
we show the annihilation cross section times the velocity for the
dark matter particle in several of these models.  The strong
velocity dependence is evident, enhancing the total cross section by
several orders of magnitude over the value at $v=0$. We show how the
low and high velocity behaviors of the cross section can be tuned
independently by varying different supersymmetric parameters. In Figure \ref{figSigvHelicity} (a) the mass of the the $A^0$ is
scanned. As it is decreased the s-channel diagram in Figure
\ref{figDiags} becomes increasingly important, the s-wave component
of the amplitude increases, and the $v=0$ cross section grows.   In
\ref{figSigvHelicity} (b) the mass of the stau is scanned.  As this
parameter is decreased the t-channel scalar exchange diagram is
enhanced and the p-wave, velocity dependent, component of the cross
section grows. Notably, we must vary the mass of the LSP along with
the stau mass in order to maintain the correct relic abundance.

The curves in  Figure \ref{figSigvHelicity} can be well fit by expressions of the form

\begin{equation}
\label{param}
\sigma v \approx (\sigma v)_0 + (\sigma v)_1 (v/c)^2
\end{equation}
where $(\sigma v)_0$ and $(\sigma v)_1$ are fit parameters.  We use
this approximation in the following sections. 

\end{section}
\begin{section}{Astrophysics}
\label{ap}
In the previous section we introduced a class of SUSY models for which the neutralino annihilation cross section shows a strong velocity dependence. In this section we explore the consequences for the density profile and the expected annihilation signal from the dark matter in the sub-parsec region around our Galactic center. To the best of our knowledge, this effect has been ignored in the literature. We find that neglecting this velocity dependence leads to an underestimation of the size of the annihilation core as well as the expected annihilation signal. In addition, we find that these corrections depend strongly on the density profile of the halo.
\begin{subsection}{Density profile}
\label{ap1}
The annihilation signal depends on the the density profile of dark matter. In particular, it is sensitive to the profile in the sub-parsec region around the central black hole where the density is expected to be quite high. Our first goal is to understand how the density profile changes in this region  when we include the velocity dependence of the cross section.

The density profile depends on a number of physical processes such as the initial phase space distribution of the dark matter particles that collapsed to form the halo, the steepening  of the profile due to the baryons, scattering by stars, loss to the central black hole, black hole or galactic merger history etc. A detailed calculation of the density profile is beyond the scope of this paper (see \cite{Bertone:2005xv} for an excellent review). We assume the following density profile \cite{Gnedin:2003rj}

\begin{equation}
\label{density}
\rho(r)=\left\{ \begin{array}{ll}
\rho(r_c) & \qquad 10r_g<r\le r_c \,\\
 \rho_0 \left({r/r_{\rm bh}}\right)^{-\gamma_{\rm sp}}  & \qquad r_c < r
    \le  r_{\rm bh} \ ,\\
 \rho_0 \left({r/r_{\rm bh}}\right)^{-\alpha} & \qquad r_{\rm bh}
    < r \ ,
\end{array} \right.
\end{equation}
where the core radius, $r_c$, is defined in the next paragraph. For $r<10r_g$ the density of dark matter particles decreases rapidly and vanishes at $r=4r_g$ \cite{Gondolo:1999ef}. In the above expression $r_g\approx 4\times 10^{-7}\, {\rm pc}$ is the Scwarzchild radius of the central black hole and $r_{\rm bh}\approx 2 \rm pc$ is the radius at which the mass of the stars within that radius is twice the mass of the central black hole. We take $\rho(r_{\rm bh})\equiv\rho_0\approx 100 M_{\odot}{\rm pc^{-3}}$, though it could be higher \cite{Gnedin:2003rj}. We parametrize our ignorance regarding the nature of the profile using the two coefficients $\alpha$ and $\gamma_{\rm sp}$. 

We now turn to the {\it core radius} $r_c$. As discussed above, the density profile is determined by self annihilation, scattering by stars, loss to the SBH etc. Scattering by stars drives the density profile to a power law \cite{Bertone:2005xv}. If the density gets too high, annihilation becomes efficient enough to prevent further rise in the density. This results in the formation of a flattened core near the galactic center. The radius at which the core starts forming is determined by 


\begin{equation}
\Gamma_{\rm ann}(r_c)\approx t^{-1}_{\rm heat}
\end{equation}
where $t_{\rm heat}\approx 2\times 10^9\,{\rm yrs}$ \cite{Spitzer:1971,Bertone:2005xv} is the timescale for heating of the dark matter particles due to scattering by stars. The annihilation rate $\Gamma_{\rm ann}(r)=\rho(r)\sigma v(r)/m$ where $m$ is the mass of the dark matter particle. The position dependence of $\sigma v$ arises due to its velocity dependence. For a virialized halo, $(v/c)^2\approx r_g/2r$). Since the dark matter density is significant for $r>10 r_g$, the relavent velocities are bounded by $(v/c)^2\lesssim0.05$.

We consider a model taken from the stau coannihilation region of mSugra.  The mass of the LSP $m=166\rm\,GeV$ and the mass of the lightest stau is $173\rm \,GeV$.
The relic density is  $\Omega_{\rm dm}h^2\approx 0.1$.  In this model,  $(\sigma v)_0=9\times10^{-30}\,\cs$ and $(\sigma v)_1=8.9\times10^{-26}\,\cs$ (see equation (\ref{param})). We will refer to this model as our fiducial model. Whenever a parameter is not explicitly defined or varied, its value is taken from this model.

For our fiducial model with $\gamma_{\rm sp}=3/2$ and $\alpha=1$, the core radius is $r_c\approx 14 r_g$, with a core density of $\rho(r_c)\approx 2\times10^8\rho_0$. If we ignore the velocity dependence, then we do not get a core. In Figure \ref{rcDensity}(a) we plot $r_c$ for different $(\sigma v)_0$ and $(\sigma v)_1$ for the same density profile. We note that the size of the core is not independent of $\sv$ and ignoring it leads to an underestimation of the core size. If $\sv$ has a dominant contribution in determining the core radius $r_c$, it has to be significantly larger than $\so$. This is because the factor $(v/c)^2\approx r_g/2r$ in front of $\sv$ is small unless we are close to the central black hole. 

Another important factor that determines the size and density of the core is the steepness of the density profile parametrized by $\gamma_{\rm sp}$. For the fiducial model, the size of the core as well as the density increases with increasing $\gamma_{\rm sp}$ as shown in Figure \ref{rcDensity}(b). The dotted lines represent the density profiles for $\sv=0$.
 \begin{figure}
 \centering
  \begin{tabular}{cc}
  \includegraphics[width=3in]{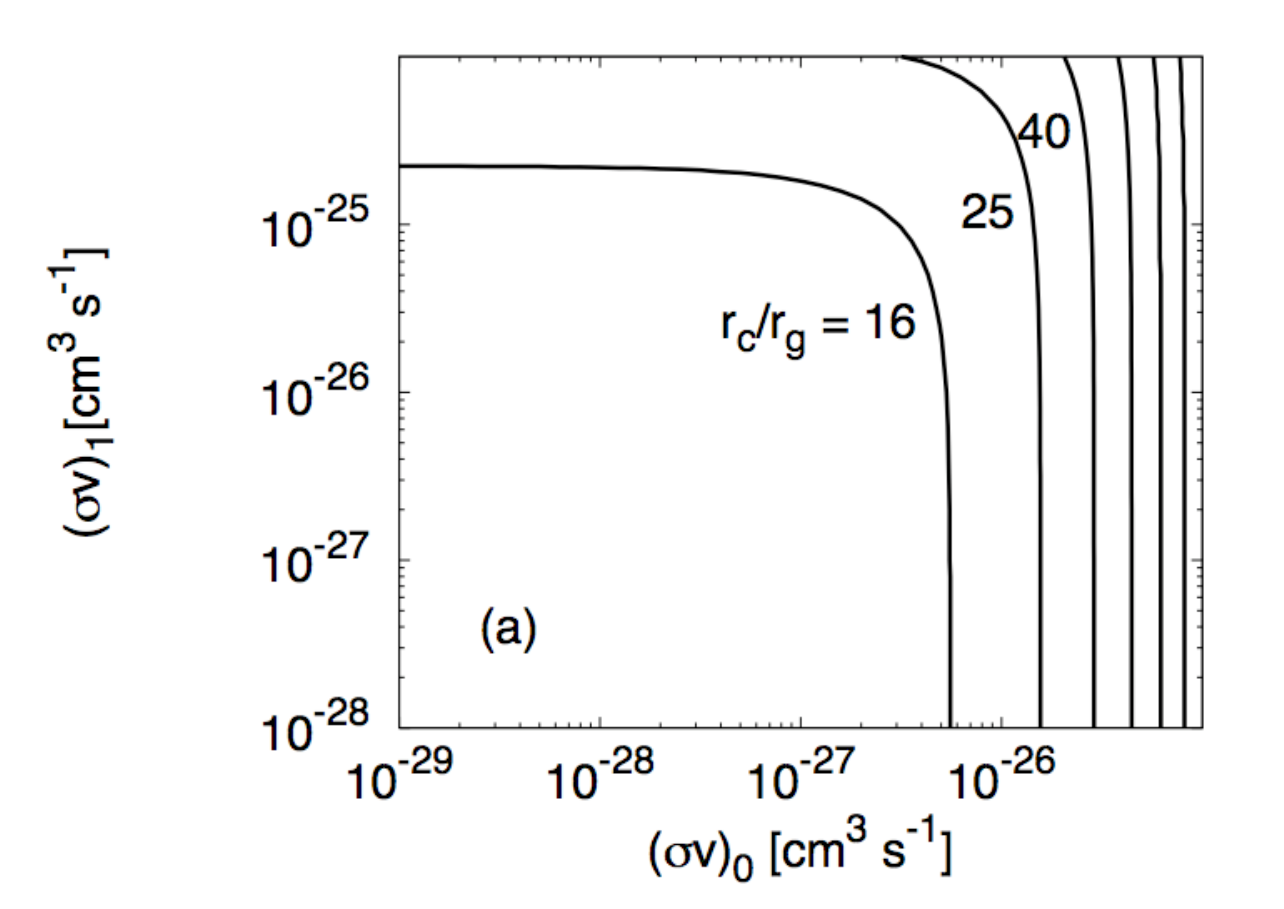}&
  \includegraphics[width=3in]{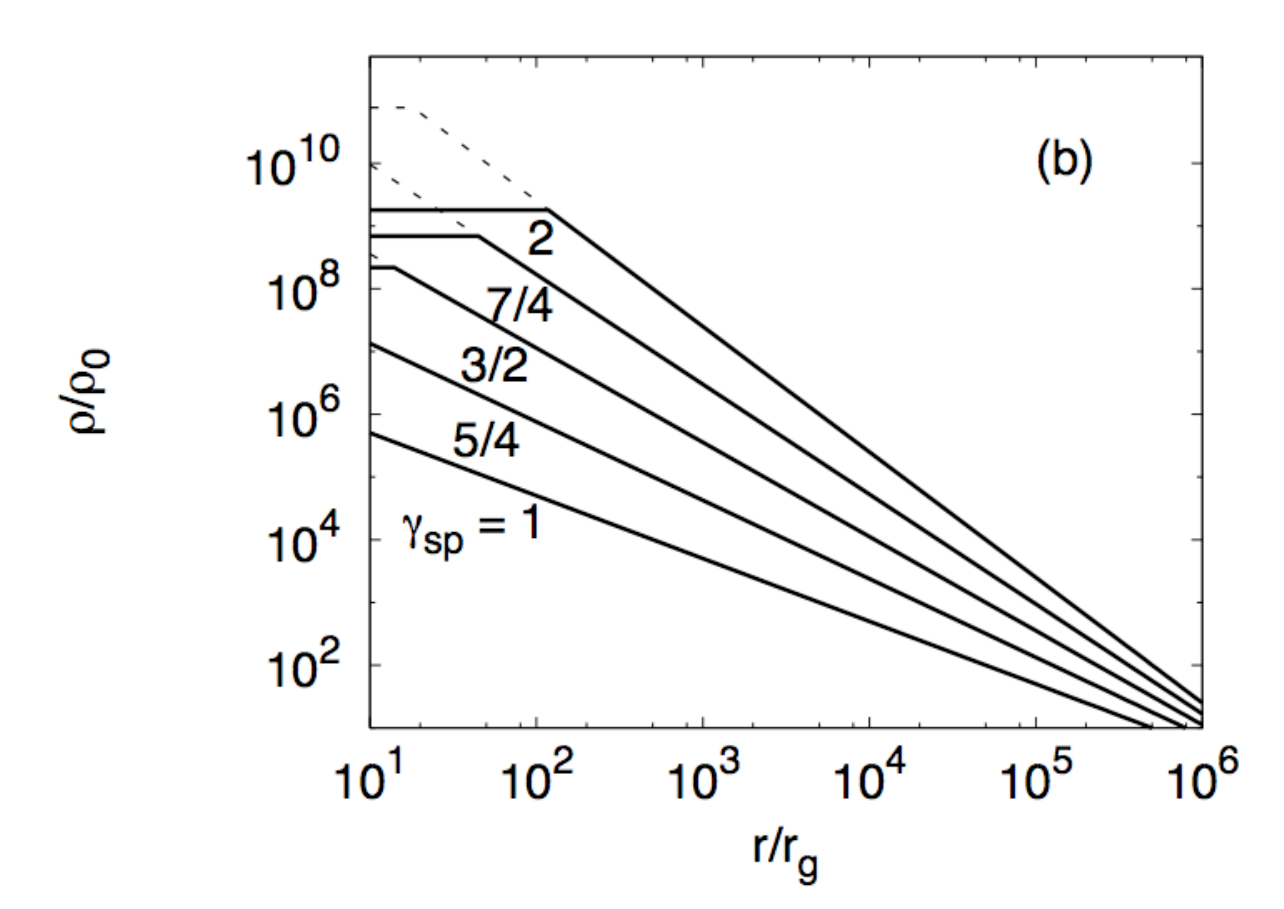}

    \end{tabular}
     \caption{(a) The variation of the core radius with $\so$ and $\sv$. Ignoring the $\sv$ leads to an underestimate of the core radius. (b) Change in the spike profile as function of $\gamma_{\rm sp}$. Note that for large values of  $\gamma_{\rm sp}$,
the radius of the annihilation core is also large. The dotted lines indicate the density profile when we set $\sv=0$. }
  \label{rcDensity}
\end{figure}

\end{subsection}
\begin{subsection}{Annihilation Flux}

The flux of photons (as observed by us) can be written as 
\begin{equation}
\label{flux}
\Phi=\frac{1}{2m^2}\int d^3{\bf r}\frac{N\sigma v({\bf r}) \rho^2({\bf r})}{4\pi|{\bf d+r}|^2}
\end{equation}
where $\rho({\bf r})$ is the dark matter density and $\sigma v({\bf r})$ is the annihilation cross section times the typical relative velocity of the annihilating particles. In the above expression ${\bf d}$ is the vector joining the sun and the galactic center, $m$ is the mass of the annihilating dark matter particles and $N$ is the number of photons (above the detector thresh-hold) produced in the annihilation process. The integral is done over a solid angle which depends on the angular resolution of the detector. We take this to be $\Delta \Omega\approx10^{-5}\rm sr$ which is the approximate angular resolution for GLAST. We remind the reader that the position dependence of $N\sigma v({\bf r})$ arises from the position dependence of the velocity in a virialized halo. Due to this position dependence of the cross section, we cannot simply separate the particle physics and astrophysics aspects of the integral as is commonly done in the literature.

The annihilation signal depends on the cross section in two ways: Explicitly through $N\sigma v$ appearing in equation (\ref{flux}) and implicity through $\rho$ which depends on $\sigma v$ as discussed in the Section \ref{ap1}.

\begin{figure}
 \centering
  \begin{tabular}{cc}
  \includegraphics[width=3in]{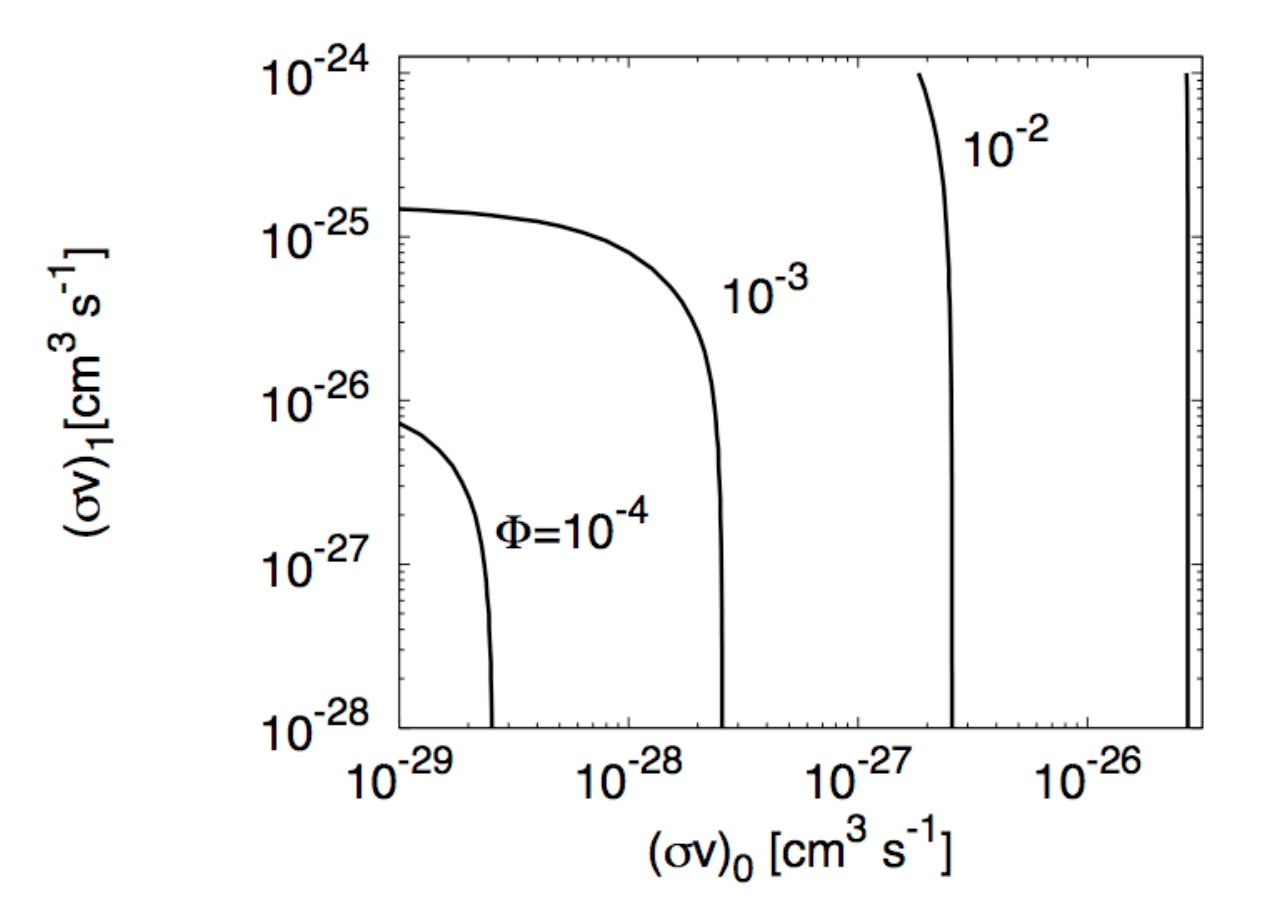}\put(-35,40){{\footnotesize{\fontfamily{phv}\selectfont (a)}}}  &
  \includegraphics[width=3in]{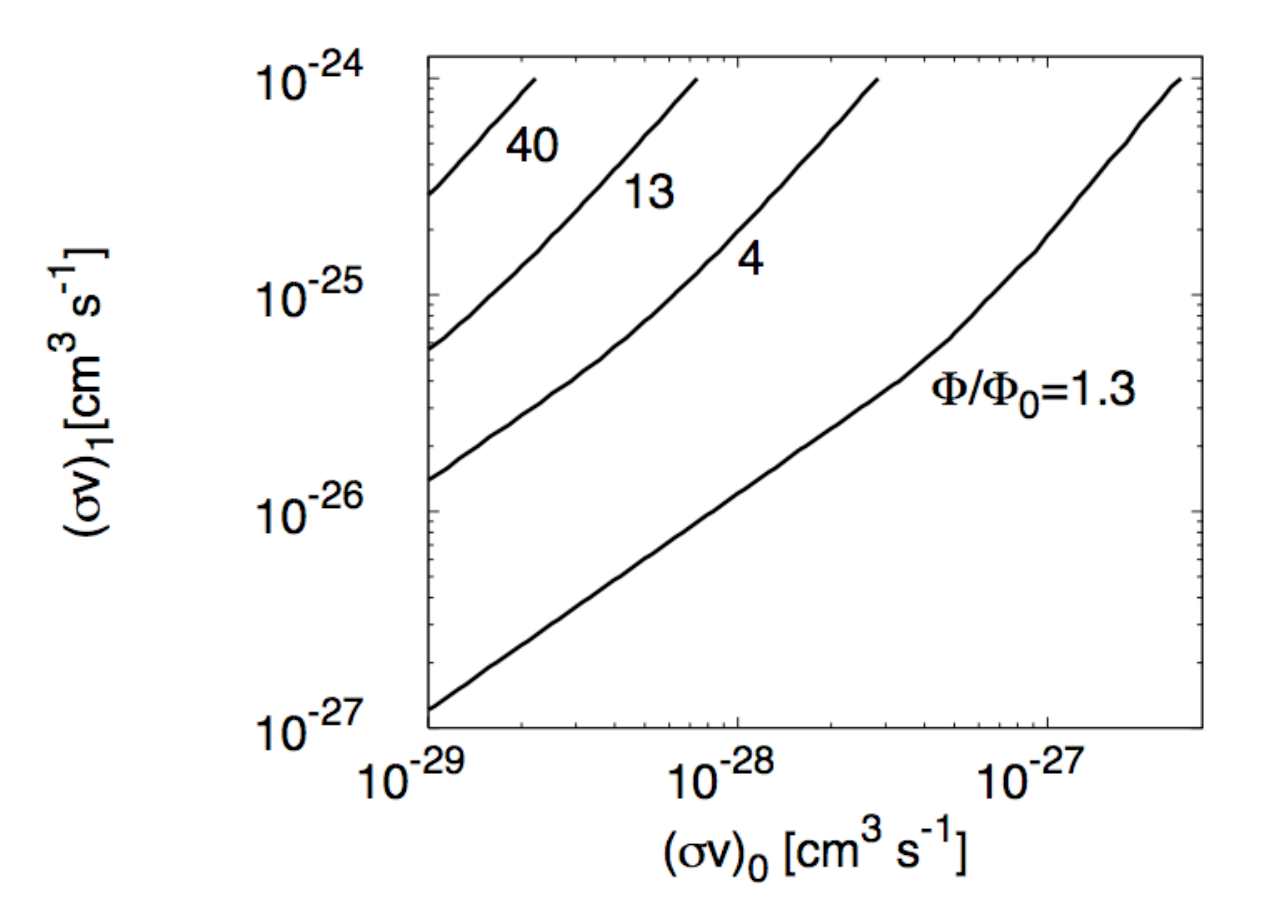}\put(-35,40){{\footnotesize{\fontfamily{phv}\selectfont (b)}}}
    \end{tabular}
     \caption{(a) The annihilation flux as a function of $\so$ and $\sv$. $\Phi$ is expressed in units of $\rm 6\times 10^{-10}cm^{-2}s^{-1}$ which is the (approximate) sensitivity of GLAST at an energy threshold of $\rm 1\,GeV$. (b) The ratio of annihilation fluxes $\Phi/\Phi_0$ where `0' refers to the flux calculated by setting $\sv=0$.}
  \label{FluxSigma}
\end{figure}

For the fiducial model with $\gamma_{\rm sp}=3/2$ and $\alpha=1$, the expected flux is enhanced by a factor of $\sim4$ compared to the case when the velocity dependence is ignored. One should view this number with caution, since it does depend  strongly on the parameters of the model.

In Figure \ref{FluxSigma}(a), we plot the annihilation flux as a function of $\so$ and $\sv$ with the same halo profile. In Figure \ref{FluxSigma}(b) we plot the ratio of the fluxes, with and without the velocity dependence in the cross section : $\Phi/\Phi_0$  where `$0$' indicates that we set $\sv=0$. As expected, ignoring the velocity dependence of the cross section leads to an underestimation of the flux. The enhancement is large when $\sv/\so$ is large.

Next,  in Figure \ref{figgamma}  we show $\Phi/\Phi_0$ as a function of $\gamma_{\rm sp}$. For $\gamma_{\rm sp}\lesssim1.6$ the enhancement increases with $\gamma_{\rm sp}$. This is due to the increase in the density very close to the blackhole compared to regions farther away which results in a greater fraction of dark matter particles having high velocities. However, as we saw in Section \ref{ap1} the size of the annihilation core also increases with $\gamma_{\rm sp}$. When the core becomes sufficiently large $(\gamma_{\rm sp}\gtrsim1.6$), the spike profile is significantly flattened (see Figure 3(a)). This decreases the enhancement of the flux.

We note that the enhancement of the signal occurs in models that are not detectable by current or planned experiments. For our fiducial model, the flux is two orders of magnitude below GLAST sensitivity (see for example \cite{Bertone:2006kr}). This is mainly due to the small $\so$ since it is $\so$ that determines the annihilation flux in regions with $r\gtrsim10^4r_g$. It is tempting to explore the SUSY parameter space with the aim of finding models with a large $\so$ and  $\sv/\so$, so that the flux is large to begin with and the velocity dependent enhancement provides a further boost. However, relic dark matter abundance constrains $\sv (v/c)^2\lesssim 10^{-26}\cs$. Thus, for $\sv/\so\gtrsim10^4$, $\so$ is typically small leading to a small overall flux.

\begin{figure}[t]
  \centering
  \begin{tabular}{cc}
  \includegraphics[width=3in]{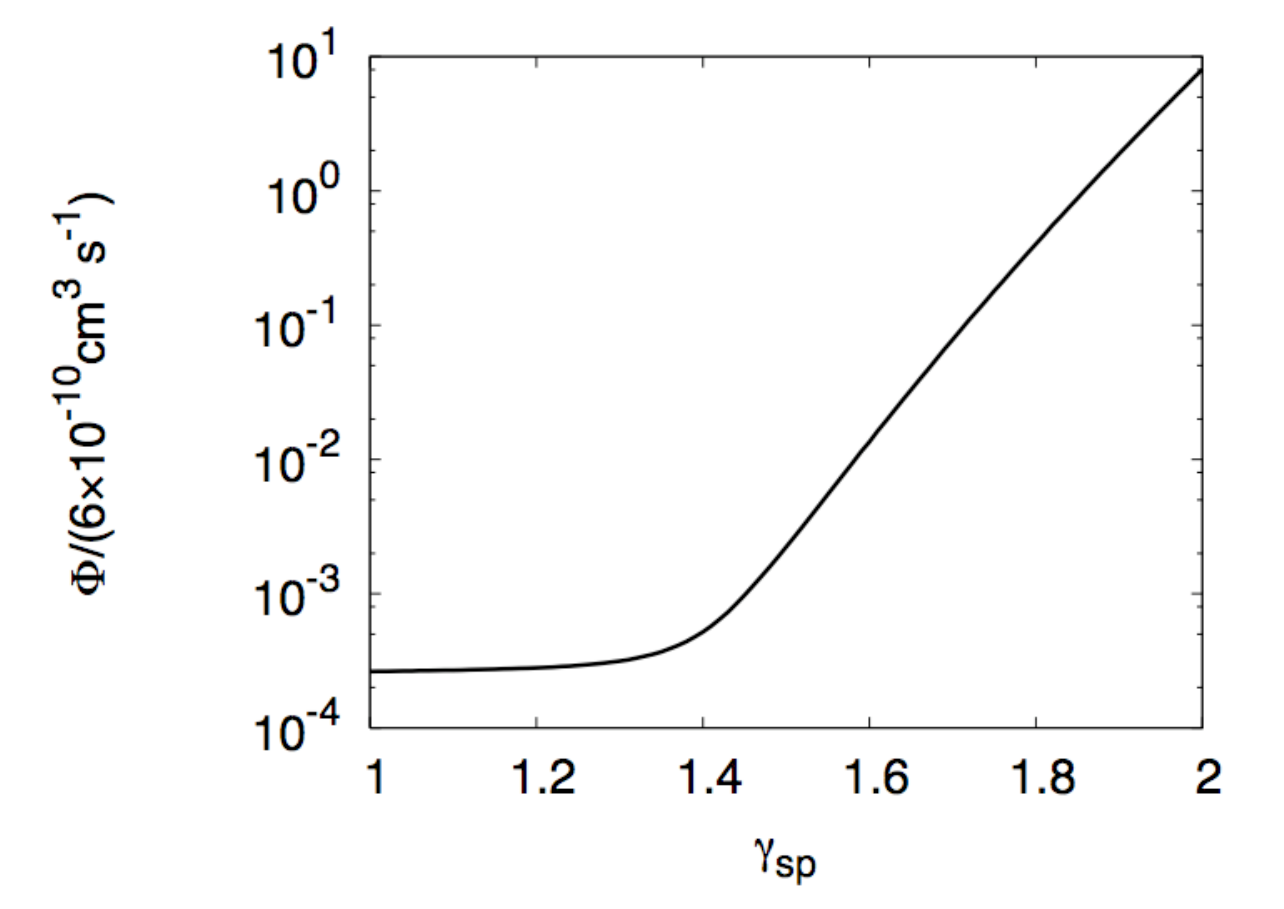}\put(-140,120){{\footnotesize{\fontfamily{phv}\selectfont (a)}}} &
  \includegraphics[width=3in]{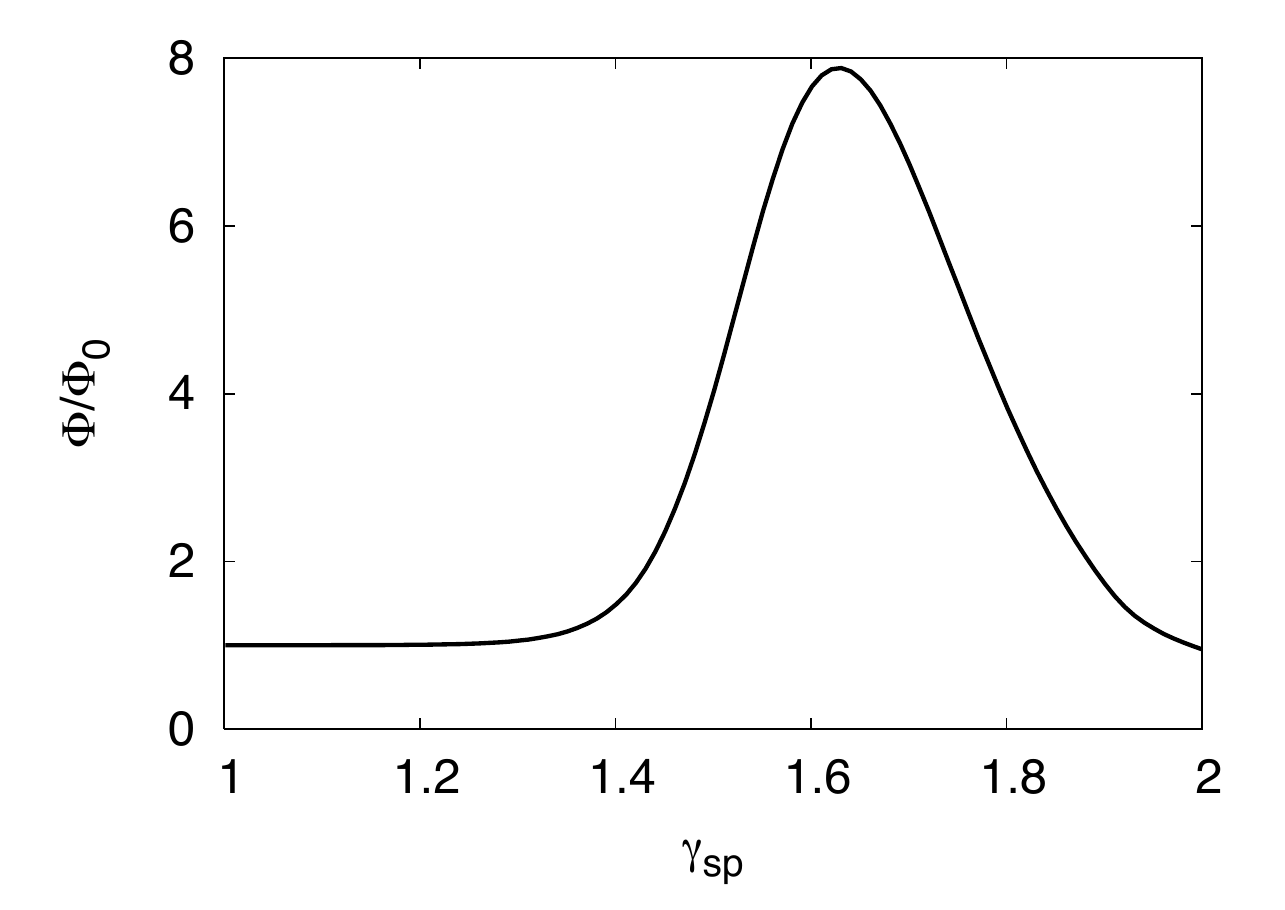}\put(-170,120){{\footnotesize{\fontfamily{phv}\selectfont (b)}}}
  \end{tabular}

  \caption{(a) Flux as a function of $\gamma_{\rm sp}$ (b) The ratio of the annihilation fluxes: $\Phi/\Phi_0$ as a function of $\gamma_{\rm sp}$. `$0$' indicates that we set $\sv=0$. The enhancement decreases after $\gamma_{\rm sp}\sim 1.6$ due to a significant increase in the size of the annihilation core.}
  \label{figgamma}
\end{figure}
\end{subsection}
\end{section}

\begin{section}{Discussion}
\label{disc}
In this paper we have discussed the consequences of relativistic dark matter
near the black hole at the center of our galaxy.  We have argued that, in
general, the commonly used approximation whereby the relative velocity of
dark matter particles is taken to vanish may be inapropriate. In regions very close to the black hole, the cold dark matter is no longer cold. If the dark matter has accumulated in a sharp spike around the black hole, this region may
account for a large fraction of the expected signal.
We presented a specific class of supersymmetric models in which
the dark matter annihilation cross section is strongly dependent on the relative velocity of the incoming particles. In these
theories, the expression for the annihilation flux no longer separates neatly into
factors depending on the astrophysics and the particle physics.
When the full velocity dependent cross section is considered, the
annihilation flux receives up to an order of magnitude enhancement over the $v=0$ value. In addition,
we found that the enhanced cross section effects the halo profile close to the galactic center.
The increased annihilations deplete the spike and widen the annihilation core.

    We explored the the change in the density profile and annihilation signal for annihilation cross sections of the form $\sigma v=\so+\sv (v/c)^2$. We showed how the annihilation core size and the flux changed as a function of $\so$ and $\sv$. To account for the astrophysical uncertainties in determining the dark matter density near the galactic centre, we presented our results for a variety of spike profiles.

None of the models we have considered are detectable by current or
upcoming gamma ray
observations.  If the neutralino is the dominant component of dark matter 
and is produced
thermally, the cross section at high velocity cannot be larger than about
$10^{-26}$ cm$^3$/sec; otherwise the relic abundance would be too small.
In most regions of the galaxy today,
the neutralino velocity $v/c$ would be very small, and the annihilation 
signal would be highly suppressed.   However, if particle physics
observations should indicate a scenario like those
we have described, it would be worthwhile to mount dedicated 
gamma ray observations concentrating on the galactic center and the 
centers of nearby galaxies.  Uniquely in those environments, in the
neighborhood of the central black holes, the annihilation cross section
would be enhance by the effect described in this paper.  

\end{section}

\begin{section}{Acknowledgments}
We would like to acknowledge Roger Blandford, Robert Wagoner, Michael Peskin, Igor Moskalenko, Peter Michelson, Elliot Bloom, Edward Baltz and Teddy Cheung for illuminating discussions. MA is supported by a Stanford Graduate Fellowship. TW is supported by the US Department of Energy, contract DE--AC02--76SF00515. 
\end{section}
\begin{section}{Appendix}

In this appendix we provide some analytic approximations to the flux integral, equation (\ref{flux}). We will assume that the velocity dependence of the cross section takes the form of equation  (\ref{param}), although this is not essential in the numerical calculations.

We split the flux integral into three parts;
$\Phi=\Phi_\textrm{core}+\Phi_\textrm{spike}+\Phi_\textrm{halo}$ bases on the density profile (\ref{density}).
In most cases, the largest contribution to the signal comes from the spike. However, the contribution from the core and halo is not always negligible. For example in the fiducial model, the spike, core and halo contribute $74, 15$ and $1$ percent of the signal respectively for an angular resolution of $\Delta \Omega=10^{-5}\rm sr$ and $\gamma_{\rm sp}=3/2, \alpha=1$. 

For the density profile, equation (\ref{density}), we can calculate the core and spike parts of the integral analytically (since $r_c,r_{\rm bh}\ll d$). For $\gamma_{\rm sp}=3/2$, the flux from the core and spike is given by (the halo integral is harder to do analytically, unless one assumes that most of the flux comes from $r\ll d$)
\begin{equation}
\begin{aligned}
&\Phi_\textrm{core}\approx\frac{1}{6}\frac{(N\sigma v)_0}{d^2}\frac{\rho^2_0 r^3_{\rm bh}}{m^2} \left[1-\left(\frac{10r_g}{r_c}\right)^3\right]\left\{1+\frac{3}{4}\frac{(\sigma v)_1}{(\sigma v)_0}\frac{r_g}{r_c}\frac{\left[1-\left(\frac{10r_g}{r_c}\right)^2\right]}{\left[1-\left(\frac{10r_g}{r_c}\right)^3\right]}\right\}\\
&\Phi_\textrm{spike}\approx\frac{(N\sigma v)_0}{2d^2}\frac{\rho^2_0 r^3_{\rm bh}}{m^2}\ln \left(\frac{r_{\rm bh}}{r_c}\right)\left\{1+\frac{1}{2}\frac{(\sigma v)_1}{(\sigma v)_0}\frac{r_g}{r_c}\frac{1}{\ln \left(\frac{r_{\rm bh}}{r_c}\right)}\right\}
\end{aligned}
\end{equation}
It is important to note that $r_c$ depends on the cross section. For the case when $\gamma_{\rm sp}\ne3/2$ we have
\begin{equation}
\begin{aligned}
&\Phi_\textrm{core}\approx\frac{1}{6}\frac{(N\sigma v)_0}{d^2}\frac{\rho^2_c r^3_c}{m^2} \left[1-\left(\frac{10r_g}{r_c}\right)^3\right]\left\{1+\frac{3}{4}\frac{(N\sigma v)_1}{(N\sigma v)_0}\frac{r_g}{r_c}\frac{\left[1-\left(\frac{10r_g}{r_c}\right)^2\right]}{\left[1-\left(\frac{10r_g}{r_c}\right)^3\right]}\right\}\\
&\Phi_\textrm{spike}\approx\frac{1}{6-4\gamma_{\rm sp}}\frac{(N\sigma v)_0}{d^2}\frac{\rho^2_0 r^3_{\rm bh}}{m^2} \left[1-\left(\frac{r_c}{r_{\rm bh}}\right)^{3-2\gamma_{\rm sp}}\right]\left\{1+\frac{3-2\gamma_{\rm sp}}{4(1-\gamma_{\rm sp})}\frac{(N\sigma v)_1}{(N\sigma v)_0}\frac{r_g}{r_{\rm bh}}\frac{\left[1-\left(\frac{r_c}{r_{\rm bh}}\right)^{2(1-\gamma_{\rm sp})}\right]}{\left[1-\left(\frac{r_c}{r_{\rm bh}}\right)^{3-2\gamma_{\rm sp}}\right]}\right\}\\
\end{aligned}
\end{equation}
Again $r_c$ and $\rho_c$ depend on the cross section and $\gamma_{\rm sp}$.

\end{section}

\end{document}